\documentclass[10pt,letterpaper]{article}
\usepackage[top=0.85in,left=2.75in,footskip=0.75in]{geometry}

\usepackage{amsmath,amssymb}
\usepackage[utf8]{inputenc} 
\usepackage{amsmath,mathtools}
\usepackage{algorithm}
\usepackage[noend]{algpseudocode}
\usepackage[colorinlistoftodos]{todonotes}
\usepackage{graphicx,graphics,subfigure,multicol} 
\usepackage{url}
\usepackage{booktabs}
\usepackage{multirow}
\usepackage{threeparttable}
\usepackage{tabularx} 
\usepackage{todonotes}
\usepackage{lscape}
\usepackage{wasysym}

\usepackage{changepage}

\usepackage{textcomp,marvosym}

\usepackage{cite}

\usepackage{nameref,hyperref}

\usepackage{microtype}
\DisableLigatures[f]{encoding = *, family = * }

\usepackage{array}

\newcolumntype{+}{!{\vrule width 2pt}}

\newlength\savedwidth

\raggedright
\usepackage[aboveskip=1pt,labelfont=bf,labelsep=period,justification=raggedright,singlelinecheck=off]{caption}

\makeatletter
\renewcommand{\@biblabel}[1]{\quad#1.}
\makeatother

\usepackage{lastpage,fancyhdr,graphicx}
\fancyheadoffset[L]{2.25in}
\fancyfootoffset[L]{2.25in}
\begin{document}
\vspace*{0.2in}

\begin{flushleft}
{\Large
\textbf\newline{Fast and principled simulations of the SIR model on temporal networks}
}
\newline
\\
Petter Holme\textsuperscript{1*}
\\
\bigskip
\textbf{1} Tokyo Tech World Research Hub Initiative (WRHI), Institute of Innovative Research, Tokyo Institute of Technology, Yokohama, Japan
\\
\bigskip
* holme@cns.pi.titech.ac.jp
\end{flushleft}

\section*{Abstract}

The Susceptible--Infectious--Recovered (SIR) model is the canonical model of epidemics of infections that make people immune upon recovery. Many of the open questions in computational epidemiology concern the underlying contact structure's impact on models like the SIR model. Temporal networks constitute a theoretical framework capable of encoding structures both in the networks of who could infect whom and when these contacts happen. In this article, we discuss the detailed assumptions behind such simulations---how to make them comparable with analytically tractable formulations of the SIR model, and at the same time, as realistic as possible. We also present a highly optimized, open-source code for this purpose and discuss all steps needed to make the program as fast as possible.


\section*{Introduction}

Infectious diseases constitute a significant burden to global health and will continue to be that for the foreseeable future. To aid policy-making, one needs to test scenarios and thus run epidemic simulations. Since such simulations rely on statistics---ideally comprising millions of simulated outbreaks---often epidemic simulations can be prohibitively slow. It is thus essential to have fast algorithms to simulate epidemics.

The standard approach for epidemic modeling is \textit{compartmental models}~\cite{hethcote2000mathematics}. These are a class of models that divide a population into different classes with respect to the disease and assign transition rules between these classes. One of the most canonical of compartmental models is the susceptible--infectious--recovered (SIR) model that assumes a scenario where people get immune upon recovery.

However, there is more to the story of how to simulate epidemic spreading than just compartmental models. The ways people come in contact so that the disease can spread is crucial for epidemics, so one should not neglect them~\cite{kiss2017mathematics}. There can be structures, or regularities, in the contact patterns that affect the disease propagation. One way of addressing this problem is to base the simulations on temporal networks~\cite{salathe2010high,masuda_holme_tnetepi,masuda_holme_f1000}. These encode who that is in contact with whom, and when these contacts happen. Temporal network epidemiology has been applied to diseases from HIV~\cite{rocha2011simulated} to influenza~\cite{salathe2010high}, from COVID-19~\cite{tnet_covid} to livestock diseases~\cite{schirdewahn2017surveillance}. It should become even more useful with the increasing availability of large-scale data set~\cite{sapiezynski2019interaction}.

Simulating the SIR model on temporal networks may seem straightforward. One can run through contacts in order of time and let each contact be a potential infection event. Still, there are many details the modeler has to sort out: How should one initialize the epidemics? How should one deal with simultaneous contacts? These seemingly technical decisions do affect the result. They may rarely affect the experiments' qualitative conclusions, but they could be large enough to hinder comparison between different studies. Thus, it is desirable for the temporal-network modelers to settle on a complete version of the SIR model. In this paper, we derived such a full model description from simple principles. There could be situations where these principles are invalid, but we believe the decisions we discuss should not be glossed over, even in such a case.

One goal for this paper is to establish an exact formulation of the SIR model in temporal networks, that could---like the Markovian SIR model on static networks---serve as a common ground for studying effects of temporal network structure, or a starting point when exploring special cases (for example what happens if the duration of infectiousness is broadly distributed). The other goal is to present a fast algorithm for simulating such a model.

This paper will proceed by discussing the principles of a full SIR model for temporal networks, then present a fast algorithm for this purpose, and finally validate and evaluate the performance of this code.

\section*{Methods}

In this section, we will discuss the considerations behind our precise formulation of the SIR model on temporal networks. We will also present and evaluate a fast algorithm to simulate this model. We assume temporal networks can be represented as a \textit{contact list}~\cite{holme2012temporal}---a list of $C$ \textit{contacts} (sometimes called ``events'') $(i,j,t)$, meaning that individuals $i$ and $j$ were in contact at time $t$. The order of the first two arguments does not matter. We will use $N$ to represent the number of nodes; $T$ to represent the duration of the data set (the time between the first and last contact). We call a pair of nodes with at least one contact an \textit{edge}.

\subsection*{Design principles}

In a broad view, it is obvious how to simulate the SIR model on contact lists: One assigns one state---susceptible (S), infectious (I), or recovered (R)---to every individual. If a susceptible appears in contact with an infectious, the susceptible can become infectious with some probability. After some time, the infectious will recover. All temporal network studies using the SIR model follow these conventions~\cite{masuda_holme_tnetepi}. However, when it comes to more subtle decisions, like how to initialize the network, there are many solutions in the literature. Some of these solutions are well-motivated, some are not. In either case, these inconsistencies between studies are undesirable.

In constructing a complete specification for the SIR model on temporal networks, we will pursue the following design principles.
\begin{description}
\item[Realism] 
If our goal is to simulate reality, the first guiding principle should be to make a realistic model. At the same time, we are willing to compromise. The ultimate purpose of this type of computational epidemiology is not accurate forecasting, but to enable researchers to compare scenarios or interventions. For example, how can one best identify influential spreaders~\cite{lee2012exploiting,starnini2013immunization} or important contacts~\cite{takaguchi2012importance}? Therefore, we will not put this principle above the others.
\item[Continuity] To compare results from temporal network studies with other representational frameworks---like static network epidemiology or differential equations---our simulations should give the same results with the same assumptions.  Static network epidemiology typically (often implicitly) assume the contacts to result from a Poisson process. This is mostly to ensure continuity to the well-mixed, Markovian SIR model---i.e., the most basic, textbook version~\cite{andersson2012stochastic,hethcote2000mathematics}. So when we run our simulation on data with exponentially distributed interevent times (like a Poisson process), then we should get the same result as with SIR on static networks.
\item[Simplicity] The ``continuity'' principle itself entails many simplifications. The point of this design principle is to keep the same level of abstraction throughout the modeling. It means that the simulation components that do not concern the limit of static network epidemiology should be as simple as those that do.
\item[Generalizability] Another criterion (which in most cases overlap with simplicity), is that it should be possible to extend the model. Relaxing one of the assumptions---say that all contacts between susceptible and infectious are equally likely to spread the disease---should not conflict with another component of the model.
\item[Speed] The above principles suffice to derive most of our detailed formulation of the model. As a tiebreaker principle, we advocate choosing options that make the simulation as fast as possible.
\end{description}

\subsection*{Precise model formulation}

As mentioned, we assume a population of $n$ individuals whose contacts are described by a contact sequence. Every individual is in precisely one state S, I, or R. If $i$ is infectious and $j$ susceptible at time $t$, then a contact $(i,j,t)$ can cause $i$ to become infectious. An infectious individual will eventually recover.

\subsubsection*{Mixed discrete and continuous times}

In most empirical temporal networks, time is discretized. For example, the widely used Sociopatterns data report contacts at 20 seconds intervals~\cite{sociopatterns}. The standard in static network epidemiology, on the other hand, is to use continuous time. This is necessary to make the model Markovian so that it reduces to the standard differential-equation version of the SIR model if the network is fully connected. Although this mix of discrete and continuous times may seem strange, it does not pose any technical or conceptual problem (cf.\ Ref.~\cite{PhysRevResearch.2.033121}). From a conceptual point of view, we can assume time is continuous---the contacts happen at integer times. It is easy, to extend the algorithm to handle floating point times. The main reason we use integer times internally is that almost all data set have times specified by integers. Note that the program still follows a continuous-time algorithm in the sense that it does not progress time step by time step.

\subsubsection*{Contagion}

We will assume that every contact represents the same probability $\beta$ of the contagion (i.e., that a contact spreads the disease). In other words, we model the contagion as a Bernoulli process on the contacts with some additional conditions---a contagion takes place at the first non-zero event after a node pair becomes SI, if the involved nodes are still SI at the time of that event.

Relative to a true epidemic situation, these assumptions are radical simplifications since many effects could cause the transmission probability to vary: The amount of pathogens emitted by different infectious individuals, or at different times by the same individual, can vary greatly~\cite{vanderwaal2016heterogeneity}. The susceptibility also varies much, not only between people but also e.g., with the time of the day~\cite{bass2016circadian,colman2018reachability}. Finally, our contact sequences do not encode the intensity of the contacts. Clearly, there is a good case for a more complex model of contagion events. The motivation we keep it this simple is the continuity principle.

In empirical data sets, nodes typically can have simultaneous contacts.  With the assertion that the contagion is instantaneous in the SIR model, simultaneous contacts become a conceptual problem. A simple solution, and the one we advocate, is to prohibit the infection from spreading further than graph-distance one per time unit of the input data. In other words, if there are contacts $(i,j,t)$ and $(j,k,t)$, but no $(i,k,t)$, then $i$ cannot infect $k$ (via $j$) at time $t$. This solution, technically speaking, makes the model a susceptible--exposed--infectious--recovered (SEIR) model where the exposed state lasts a time less than the data's resolution. However, it simplifies the code a great deal and probably makes the simulation more realistic (since SEIR is always more accurate than SIR). Moreover, once again, except for some extreme cases, this decision will not significantly affect the output.

A different principle one could potentially follow would be to assume that the contacts happen at different (continuous) times but that these times have been truncated to integers. Then the principled approach would be to sample the contacts of nominally the same time in random order and average over different realizations of this random sampling~\cite{holme2015information}. This approach will make more sense if the data set's time resolution is relatively low compared to the propagation of the disease. However, in such a case, one should probably instead consider a static network model since the temporal information would be less critical.

\subsubsection*{Recovery}

For the time to recovery $\delta\geq0$, we simply follow the standard Markovian SIR model for static networks and sample it from an exponential distribution
\begin{equation}\label{eq:intereventtime}
   \text{Prob} (\delta)= \nu\exp(-\nu\delta) .
\end{equation}
If one represents times as integers internally, one needs to round the sampled times down to the nearest integer.

The duration of real infections is typically not exponentially distributed~\cite{vergu2010impact}, so this is not a choice made for realism, but to conform to the standard in static network epidemiology. If one would want the recovery times to follow a particular distribution other than exponential, there is no problem to just replace the exponential random numbers when obtaining a recovery time. (This is unlike static network epidemiology where using a different distribution of recovery rate demands a different algorithm.~\cite{masuda2018gillespie})

Some papers use a fixed time for the infection duration~\cite{holme2015information,lee2012exploiting}. This does not simplify anything, is it probably not more realistic, as realistic distributions of infectiousness tend to be peaked and skewed~\cite{krylova2013effects}. Furthermore, they could cause unrealistic threshold effects (when a gap in the contacts is very close to $\delta$). So since there are no major advantages with this approach, exponentially distributed infection times must be preferable.

\subsubsection*{Number of sources}

For several reasons, we recommend starting the outbreak at one node, rather than many. The main reason for this is that medical epidemiology is usually concerned with the outbreak of one pathogen or entering the population, typically via one external (zoonotic) interaction, or arising from a mutation in one host. Starting the epidemics at different sources, one would be assuming that there had been some spreading outside of the considered network. In other words, that one is modeling an open system. In that case, one would also need to model the influx of pathogens during the outbreak, which adds another level of complexity to the problem.

Another conceptual problem with having many sources is how to choose them. Unfortunately, there is no rationale to follow that is both simple and consistent with fundamental epidemiological facts. If one, for example, is modeling bioterrorism, having many seeds could make sense. However, then a modeler needs to know whether the adversary chooses seeds to optimize the damage~\cite{jankowski2018probing}, just at random or in some cluster of the network.

Furthermore, and maybe most importantly, by using many sources, one misses the early die-outs characteristic of epidemic models~\cite{holme2015information} and presumably also real epidemics. An outbreak typically either dies very early or takes off to follow a predictable curve~\cite{janson2014law}. With several sources, the early die-offs become inaccessible.

Finally---and related to the previous point---with only one seed, one can measure the basic reproductive number $R_0$ directly~\cite{holme_masuda}. This is one of the most fundamental epidemiological quantities defined as the expected number of others that the source will infect. Note that one cannot avoid stochastic simulations to calculate this number because neighbors of the source could get infected by other nodes than the source and would not contribute to $R_0$.

\subsubsection*{Initialization}

Now that we established the need for only one infection source, then how should we choose it? In the spirit of simplicity, we chose it with uniform randomness. This is also related to realism---there might be some correlation between network positions and the chance of acquiring a zoonotic infection. However, without additional knowledge, we cannot do better than choosing it randomly.

In the spirit of simplicity, we also choose the time of the infection uniformly at random between the beginning and end of the contact data set. Introducing the infection at a time related to features of the data---like the beginning of the data or when the seed enters the data---could introduce biases. Since the disease enters from the outside by a process unrelated to the SIR dynamics, we should randomly choose the time. Of course, there is a chance that the outbreak will start toward the end of the data set and thus not have enough time to spread far. Therefore optionally, one could choose the starting time randomly in an early time interval. Nevertheless, there is no simple rule to chose that interval. If the research purpose is to investigate the largest possible outbreaks, one must find such a rule, even though it has undesired consequences. Otherwise, we recommend picking an infection time by uniform randomness in $[0,T)$. For the rest of the paper, we will follow that principle.

\subsubsection*{Summary}

Summarizing the above points, a precise formulation of the SIR model on contact sequences is as follows.
\begin{description}
    \item[Initialization] Initialize all individuals to susceptible.
    \item[Seeding] Pick a random individual $i$ and a random time $t_i$ in the interval $[0,T)$. At time $t_i$, infect $i$.
    \item[Recovery] Whenever a node becomes infected, let it stay infected for an exponentially distributed time $\delta$ before it recovers.
    \item[Contagion] If $i$ got infected at time $t_i$ and is still infected at time $t>t_i$, and $j$ is susceptible at time $t$, then a contact $(i,j,t)$ will infect $j$ with probability $\beta$.
\end{description}

\subsection*{Algorithm}

Now we describe the algorithm. The code, written in C and Python, is available at \url{github.com/pholme/tsir/}. 
This code is commented and written for clarity. Thus, we prioritize to describe the ideas rather than all the algorithmic details. We recommend the reader follow the actual code when reading this section.

\subsubsection*{Straightforward algorithm}

The simplest way of simulating the SIR model on a temporal network is to:
\begin{enumerate}
    \item Initialize all nodes as susceptible.
    \item Run through the contacts in increasing order of time.
    \item If a there is a contact between a susceptible and infectious node, then make infect the susceptible node with probability $\beta$.
    \item Whenever a node gets infected (including the source), then draw its time to recovery from an exponential distribution, and change its state to I.
    \item Stop the simulation when there are no infectious nodes.
\end{enumerate}
There are many tricks to speed up such a simulation. For example, one can use bisection search to find the first contact capable of spreading the disease (and thereby avoid scanning through all contacts before introducing the infection). Another trick is to note when individuals become inactive and stop the simulations when there are no active contacts. Still, the running time of this algorithm (above the epidemic threshold) will be linear in $C$. Now, consider the contacts between a pair of nodes. There could be thousands of these, but only one of them can spread the infection. So clearly, if we can identify that particular contact without having to scan through all contacts, that could make the algorithm much faster for dense enough data sets.

\begin{figure}
\includegraphics[width=\columnwidth]{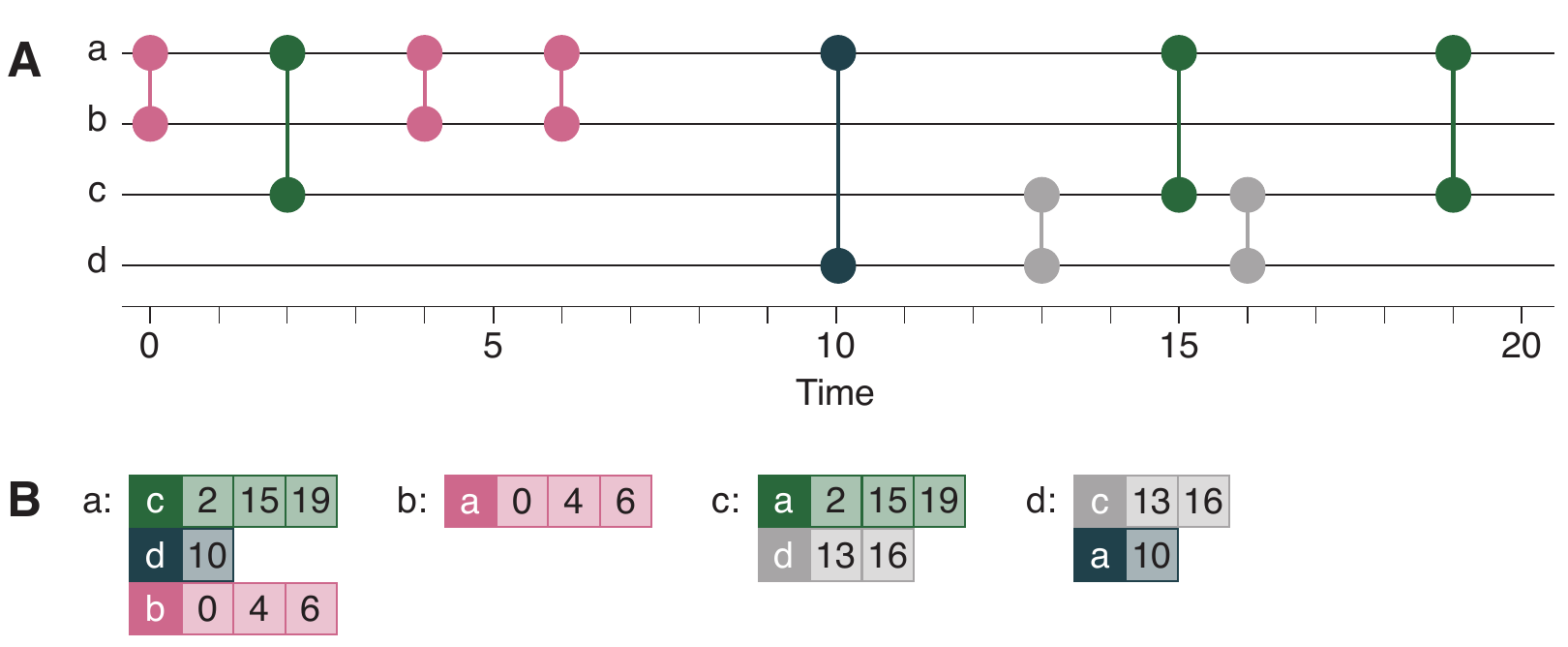}
\caption{\textbf{Internal representation of the temporal network.} In this figure, panel B illustrates how the temporal network in panel A is represented internally. In A, we display a temporal network of four nodes $({\rm a}, {\rm b}, {\rm c}, {\rm  d})$, four edges ($({\rm a},{\rm b})$, $({\rm a},{\rm c})$, $({\rm a},{\rm d})$, $({\rm c},{\rm d})$) and nine contacts ($({\rm a},{\rm b},0)$, $({\rm a},{\rm c},2)$, etc.). In panel B, the internal representation is organized by nodes. Every node has a list of neighbors (e.g., node a has the neighbor list $({\rm c}, {\rm d}, {\rm b})$). For every neighbor, there is an ordered list of times of contacts with that neighbor (e.g.\ a has contact with c at times $(2,15,19)$). The neighbor lists are ordered in decreasing order of the last contact with that neighbor. This makes it possible to break iterations over neighbors if the infection time of a node is later than the last contact with the neighbor.
}
\label{fig:internal}
\end{figure}

\subsubsection*{Event-based algorithm}

The faster algorithm that we will discuss is inspired by the \textit{event-driven algorithm} for SIR on static networks by Kiss, Miller and Simon~\cite{kiss2017mathematics}.

To understand our algorithm, first consider a pair of nodes $(i,j)$ and assume one of them, say $i$, gets infected at time $t_i$. Now assume no other nodes can infect $j$ other than $i$. The infection process between $i$ and $j$ is then a Bernoulli process of a finite number of binary random variables with probability $\beta$. (Note that the corresponding part of the event-based algorithm for SIR on static networks is a Poisson process.) The number of Bernoulli random variables is the number $n_{ij}(t_i)$ of contacts between $i$ and $j$ for $t>t_i$. The probability that the $k$'th such contact will transmit the disease is given by
\begin{equation}
    \beta(1-\beta)^{k-1} .
\end{equation}
One can sample such a random number $k$ by
\begin{equation}\label{eq:bernoulli}
\left\lfloor\frac{\log (1-X)}{\log (1-\beta)}\right\rfloor .
\end{equation}
where $X$ is a standard, uniformly distributed random variable $X$ on the unit interval $[0,1)$. Note that the above operations take $\mathrm{O}(\log c)$ time (for a list of $c$ contacts between two nodes), compared to linear time for just scanning through the contacts. 

Using the above strategy, when a node $i$ gets infected, we can go through its neighbors $j\in\Gamma_i$, and calculate if $j$ could be infected by $i$, then which one of the contacts between $i$ and $j$ would transmit the disease. In the C code, this happens by calling a subroutine, \texttt{contagious-contact}, that takes $i$'s infection time $t_i$ and the time-ordered list of contacts between $i$ and $j$, $\mathbf{t}_{ij}$, as input. Then proceed as follows:
\begin{enumerate}
\item Use bisection search to find the smallest index $k$ of $\mathbf{t}_{ij}$ such that $t_i<\mathbf{t}_{ij}(k)$. Where $\mathbf{t}_{ij}(k)$ denotes the $k$'th contact of $\mathbf{t}_{ij}$.
\item Add a random number $K$ generated by Eq.~\ref{eq:bernoulli} to $k$ and call it $k'$.
\item If $k'$ is larger than $\mathbf{t}_{ij}(k)$'s number of elements, then return some out-of-bounds value (to signal that no contact will spread the disease). Otherwise, return $k'$---the contact between $i$ and $j$ that could be contagious.
\end{enumerate}

From the previous section, we can see that our code needs a priority queue---a data structure where one can quickly delete the smallest element and insert arbitrary elements. There are many ways to implement a priority queue. For our situation (where we have to delete, update, and add elements), operating a priority queue of length $n$ has at least a complexity of $\mathrm{O}(\log n)$. Among algorithms with this complexity, we use perhaps the simplest one---a \textit{binary heap}. Apart from its simplicity, one appealing feature of using a binary heap for this problem is that updating the entry for an infected node already on the heap is very fast. Briefly speaking, updating a heap needs two types of sorting operations---\texttt{heap-up} and \texttt{heap-down}---where \texttt{heap-up} is much faster, and the only one needed to update elements already in the heap. We only use \texttt{heap-down} when we delete the smallest element.

The core of the code happens in a subroutine called \texttt{infect} that handles the infection of one node:
\begin{enumerate}
\item Pop the individual $i$ with the earliest infection time from the heap.
\item\label{step:get_nb} Iterate through the neighbors $j$ of $i$.
\begin{enumerate}
\item If $j$ is susceptible, get the time $t_j$ when it would be infected by $i$ (by calling \texttt{contagious-contact}).
\item If there is no earlier infection event of $j$ on the heap, or $i$'s recovery time is earlier than $t_j$, then put the contagion ($i$ infects $j$ at time $t_j$ on the heap).
\end{enumerate}
\end{enumerate}
A trick to speed the code up is to sort the neighbor list in decreasing order of time of the last contact---see Fig~\ref{fig:internal}. In that way, we can break the iterations over neighbors (step~\ref{step:get_nb}) whenever we encounter a neighbor with which $i$ has no future contacts.)

Then the final structure of the program is simply:
\begin{enumerate}
\item Read the network and initialize everything.
\item\label{step:source} Infect the source node.
\item While there are any nodes left on the heap, call \texttt{infect}.
\item Reset the simulation.
\item Go to \ref{step:source} until you have enough averages.
\item Evaluate the output.
\end{enumerate}

\begin{figure}
\includegraphics[width=.7\columnwidth]{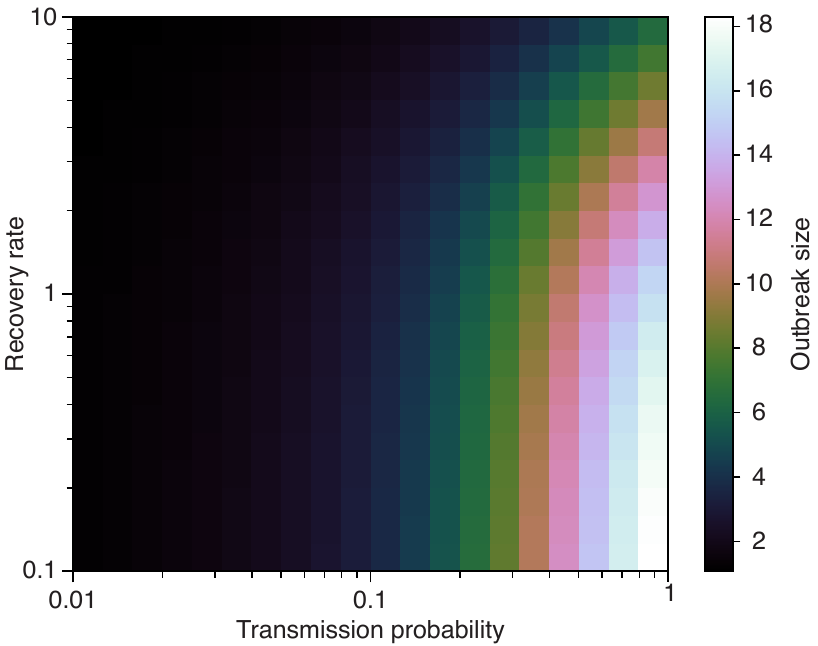}
\caption{\textbf{Example output.} This heatmap shows the average outbreak size as a function of the model parameters. The raw data comes from the first day of sampling in Ref.~\cite{Wouter-Van-den-Broeck:2012fk}. It represents the proximity patterns of visitors to an art gallery.
}
\label{fig:gallery01}
\end{figure}

\subsubsection*{Further notes about the implementation}

Our implementation of the above algorithm, available at \url{github.com/pholme/tsir/}, uses a mix of C and Python. The idea is to exploit C's speed for the core routines and the many libraries of Python to simplify the pre- and post-processing. We have not made this into a full Python library because research building on this codebase would most likely need to add functionality on a low level. We have refrained from adding many imaginable measurements, both because it is hard to envision a sufficiently complete list of such, and it would slow down the program. We display an example output of the program in Fig~\ref{fig:gallery01}.

We use a 64-bit state version of the PCG (Permuted Congruential Generator) random number generator~\cite{oneill:pcg2014}. For this type of simulation, neither speed nor statistical quality of the random number generation is critical. For simplicity, we could just have used some lower-performance, library generator. Still, in the spirit of using state-of-the-art components, we opt for PCG. For some parameter values, it does save a few percent of computing time compared to popular random number generators of the previous generation (i.e., the Mersenne Twister).

\begin{figure}
\includegraphics[width=\columnwidth]{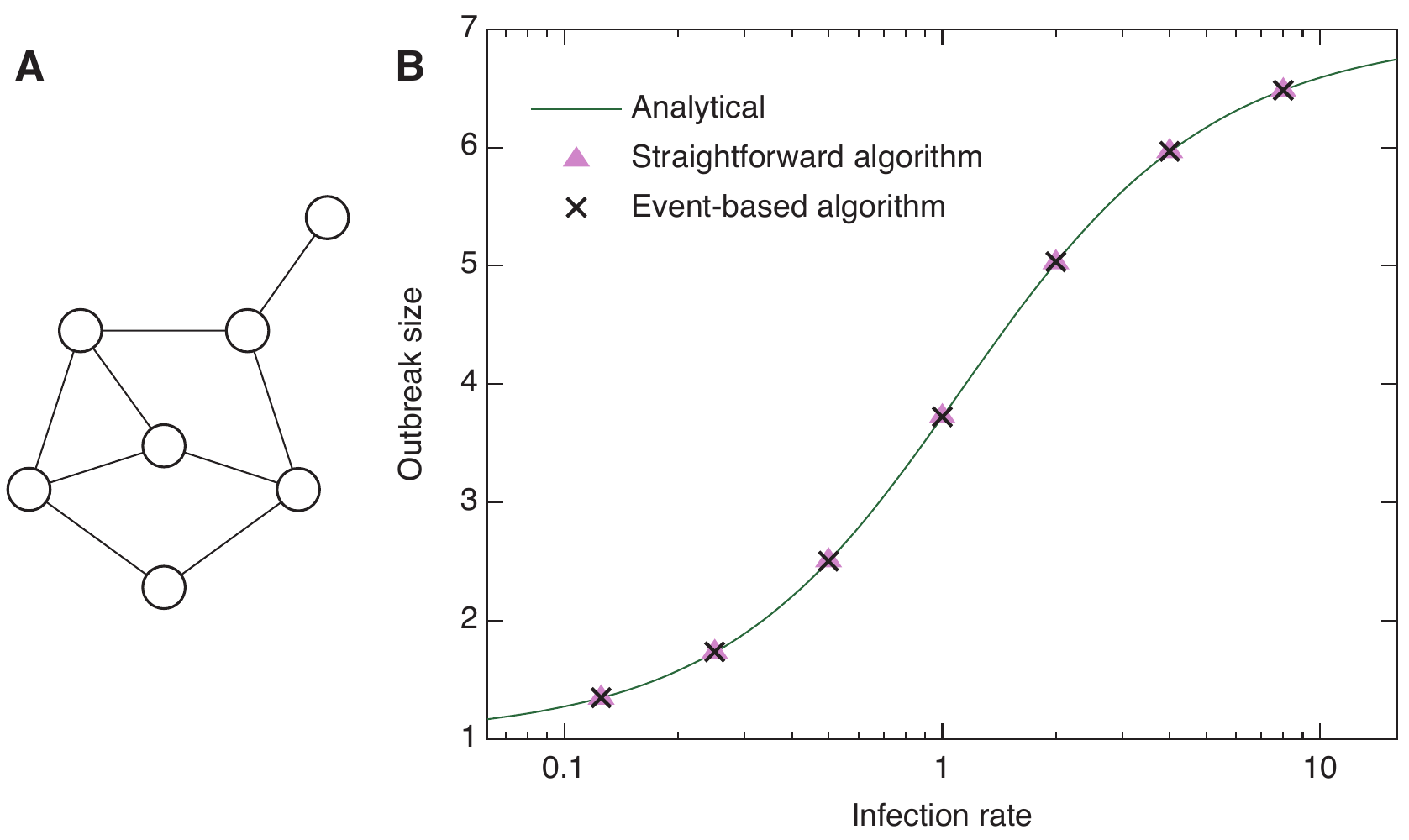}
\caption{\textbf{Validation of the program.} Panel A shows a small graph with especially complex behavior with respect to the SIR model (and thus a good test case). Panel B shows the predicted outbreak size for the graph in A. The solid curve is the analytical solution. The symbols represent averages over $10^6$ values for the straightforward and event-based algorithms, respectively.
}
\label{fig:special}
\end{figure}

\begin{figure}
\includegraphics[width=0.7\columnwidth]{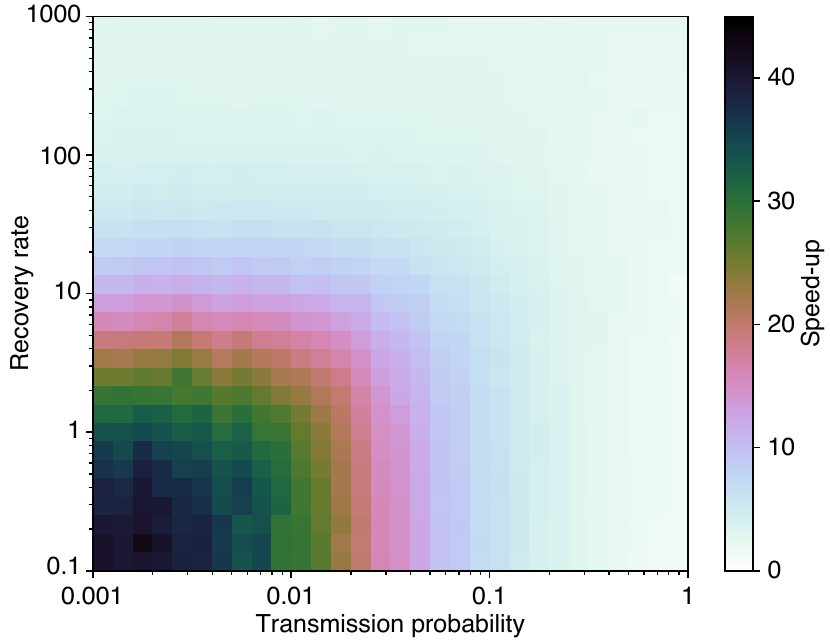}
\caption{\textbf{Speed-up relative to the straightforward algorithm for an artificial network as a function of the SIR model parameters.} How many times faster the event-driven program is compared to the reference code for the same data set as in Fig~\ref{fig:gallery01}. The minimum value of the speed-up in this figure is $2.8$.
}
\label{fig:gallery_speedup}
\end{figure}

\begin{figure}
\includegraphics[width=.9\columnwidth]{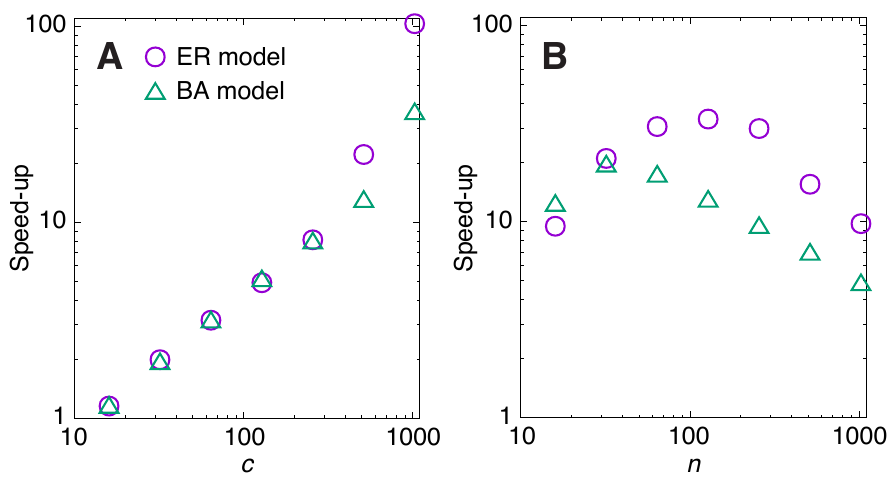}
\caption{\textbf{Speed-up relative to the straightforward algorithm of artificial networks.} How many times faster the event-driven program is compared to the reference code. The underlying temporal networks are Erd\H{o}s-R\'enyi---ER, binomial random graphs, $G(n,z/n)$ (where $z$ is the average degree---for Barab\'asi-Albert (BA) model networks~\cite{mejn:book}. We add exponential inter-event times to these static graphs ($\lambda=1/c$) added until $T$ is at least one. (Where $\lambda$ is the usual, ``rate parameter,'' of the exponential distribution.) We use $10^3$ temporal networks and $10^6$ outbreak runs per set of parameter values. The recovery rate is $\nu=1$. Error bars (standard errors) would have been smaller than size of the symbols and thus not shown. Panel A shows the scaling of the speed-up as a function of the average number of contacts per link. Here $n=128$ and $z=1024/c$ (to keep the number of contacts constant). Panel B displays the speed-up as a function of $n$. Here $c=512$ and $z=2$.
}
\label{fig:real_scaling}
\end{figure}

\section*{Results}

In this section, we go over some analysis of our temporal-network simulation program.

\subsubsection*{Validation}

We validate our event-based program by consistency checks in several ways. In this section, we will discuss some such checks---validation against the straightforward implementation and the analytical solution of the standard Markovian SIR model on static networks. As mentioned above, if the contacts are generated by a Poisson process on the edges---i.e., if they have exponentially distributed times between the contacts---then SIR on a temporal network and static network simulations should give the same results.

For this test, we use the graph shown in Fig~\ref{fig:special}A. This graph has a complex behavior with respect to the SIR model. It is also small enough to solve exactly~\cite{holme2017three}---the expected outbreak size $\Omega$ as a function of $\beta$ is
\begin{equation}\label{eq:special}
\Omega(\beta)=\frac{1176\beta^{10}+8540\beta^9+\dots+123\beta+7}{168\beta^{10}+1316\beta^9+\dots+105\beta+7} .
\end{equation}
We add contacts to this graph's edges by drawing exponential random numbers with the rate parameter one. We break these time series when they are longer than $\tau=1000$ (arbitrary units). This gives an expected number of 1000 contacts per edge. Since our code reads integer time stamps, and we want as high resolution as possible, we rescale the times so that $T=2^{32}-1$ which guarantees there is no integer overflow. Note that there is a trade-off: If we have too many contacts per link, then the distribution of inter-event times gets further from exponential. If we have too few contacts per link, there is a higher chance the outbreak will not die by the end of the data set. We use $\nu=1$ and average over $10^6$ runs of the algorithms and $100$ realizations of the generation of time-stamps. Even with these caveats, $\Omega$ from the temporal network simulation is statistically indistinguishable from the exact values from the static network version (standard scores between $0.5$ and $2$). Furthermore, the straightforward algorithm and the event-based algorithm are also indistinguishable.

\subsubsection*{Time complexity}

The worst-case complexity of the event-driven algorithm is $O(n^2\log n\log C)$ for a dense network (where $C$ is the number of contacts). Each node has to enter and exit the priority queue (a factor $n\log n$). For an infected node, all the neighbors need to be scanned (another factor $n$). Then for each neighbor, the infecting contact needs to be identified (in a worst case this has a complexity $\log C$). Most real networks of interest are sparse---i.e.\
the degrees are bounded, giving the complexity $O(n\log n\log C)$---and has $C\gg n$.

The straightforward implementation is $O(C+n)$ in a worst case, and should be slower as long as $C$ is sufficiently large. Conversely one could construct temporal networks where the straightforward algorithm is faster---make a list of one contact per node pair for all node pairs, then repeat the same list after the first. Is such a temporal network the number of links nodes is maximal and the number of contacts per link is low. Furthermore, every node is reachable from every other. This should be a case where the straightforward algorithm outperforms the event-based one. However, empirical temporal networks typically look very different with very large $C$ values and relative low reachabilities---i.e.\ many node pairs are unreachable due to the constraint that paths have to follow increasing time stamps reduces the outbreak sizes, so that a large part of the network will never be reached even in a worst case. In the event-based simulations, the program does not need to evaluate contacts in these unreachable parts of the networks, which contributes to its speed in practice. Many other factors affect the running time. For example, the earlier the outbreaks die, or the smaller they get, the shorter are the execution times. For these reasons, it is challenging to make a complete theory of these algorithms' relative running times for practical parameter values.

\subsubsection*{Evaluation}

To evaluate the speed of our event-based algorithm, we use artificial temporal networks. We generate these in a similar way to the ones used to check the limit to the static network SIR model described in the previous section. The difference is that we here use random graphs---the standard Erd\H{o}s-R\'enyi or $G(n,p)$ model~\cite{mejn:book}---and Barab\'asi-Albert models as the underlying structure. Then we put time series of contacts, with inter-event times drawn from an exponential distribution, on the edges.

We compare our event-driven algorithm to the straightforward method. For a fair comparison, we employ all simple optimizations that we can think of for both programs---such as bisection search to find the earliest contact after the beginning of the epidemics. We report the times of the disease simulation, i.e.\ excluding the time to read the data and fill up the data structures (which is not the bottleneck of the program).

In our first experiment---see Fig~\ref{fig:gallery_speedup}---we check the relative speed-up for the same data set as in Fig~\ref{fig:gallery01}. We note that the event-based algorithm always outperforms the straightforward one, although the region of the parameter space where the speed-up is larger than around 3 is not that large. This region is at small transmission probabilities and small recover rates, i.e.\ the disease does not spread much, still it does not die out. In this case, the straightforward algorithm still has to go through all contacts, whereas the event-driven algorithm just has to go through the few ones that get infected.

In our second experiment, we use $10^3$ temporal networks for averages, and $10^6$ outbreak simulations per network. We chose the parameter values in such a way that the outbreak sizes should be intermediate. In Fig~\ref{fig:real_scaling}, we show the speed-up---the execution time of the straightforward implementation divided by the time of the event-driven simulation. We ran the simulations on a workstation with dual AMD EPYC 7552 CPUs, 256Gb RAM memory and 192 logical cores (at least half were idle during the experiment).

As predicted, more contacts per edge increases the advantage (Fig~\ref{fig:real_scaling}A). To keep the total number of contact the same, we let the average degree be $z=1024/c$. This means that for the largest values of $c$ of the Erd\H{o}s-R\'enyi model, the networks are fragmented, which explains the increase in the speed-up relative to the BA model (which never is fragmented). For fragmented networks, the event-based algorithm never needs to deal with connected components other than the one where the disease starts, which contributes to its speed.

In Fig~\ref{fig:real_scaling}B we see that the speed-up has intermediate peaks both for the Erd\H{o}s-R\'enyi  and the Barab\'asi-Albert models. Why the relative speed-up decays for larger $n$ is hard to say. The most important message from the results of this study is that the event-driven algorithm is always faster than the straightforward one. It is also important to notice that the practical run times cannot be explained by any single parameter, not even output quantities like the average outbreak size or time to extinction. Rather the practical run times depend both on the progression of the simulated outbreak and properties of the network that do not affect the disease spreading.

The code used in this experiment is available here \url{https://github.com/pholme/tsir_eval/}.

\section*{Discussion and Conclusion}

We have derived a principled detailed formulation of the SIR model on temporal networks and presented a fast, open-source simulation code for this model.

The event-based algorithm that we proposed can be extended to many other compartmental models. As long as individuals do not reenter the susceptible state, it should be quite straightforward to extend our code. This would cover e.g., the SEIR model~\cite{salathe2010high}. The only significant difference would be that one needs to put different types of events in the priority queue. For models like the SIS, where individuals can become susceptible again, it might be hard to write efficient event-driven code. For example, in this case, one can no longer discard potential infection events because they happen later than others. Probably other ideas for fast epidemic simulations could work better in this case~\cite{STONGE201930}.

Another direction for future research would be to sacrifice some of the principles that we recommended to further increase the speed. One could, for example, use a fixed duration of the infectious stage~\cite{lee2012exploiting}. In such a case, the gaps between the contacts determine whether an edge could transfer the infection. This, we speculate, could open for other types of fast algorithms. In general, temporal networks opens many intriguing problems for algorithm design. We recommend Ref.~\cite{arash,himmel,petrovic2019counting} for further inspiration.

\subsubsection*{Acknowledgements}

We thank Gordon Erlebacher and Martin Sterchi for constructive comments.

\bibliography{tsir}

\end{document}